\documentclass[fleqn,twoside]{article}
\usepackage{gc}
\usepackage[english,russian]{babel}
\usepackage[T2A]{fontenc}

\heads{G.S. Bisnovatyi-Kogan and O.Yu. Tsupko}
      {PRIMORDIAL BLACK HOLE: MASS AND ANGULAR MOMENTUM EVOLUTION}

\begin{document}
\twocolumn[ \Arthead{00}{2008}{0 (00)}{0}{00}

\Title{PRIMORDIAL BLACK HOLE:        \yy
       MASS AND ANGULAR MOMENTUM EVOLUTION}

\textbf{G.S. Bisnovatyi-Kogan}$^1$$^*$$\dag$$\ddag$ \textbf{and
O.Yu. Tsupko} $^2$$^*$$\ddag$ \\

\textit{$^*$Space Research Institute of Russian Academy of Science, Profsoyuznaya 84/32, Moscow 117997, Russia\\
$\dag$Joint Institute Nuclear Research, Dubna, Russia\\
$\ddag$Moscow Engineering Physics Institute, Moscow, Russia}
                %%   different addresses +)
              %%
%%%
%%%  +  In more complex cases please give your information
%%%     in an arbitrary (but unambiguous!) form.
%%%

\Abstract
    {The evolution of the primordial low mass black
    hole (PBH) in hot universe is considered. Increase of mass and
    decrease of PBH spin due to the accretion of radiation dominated matter
    are estimated with using of results of numerical simulation of PBH
    formation and approximate relations for accretion to a rotating black hole.$^3$}

]  %%%%%%%%%%%%%%%%%%%%%%%%%%%%%  End of temporary one-column mode
\email 1 {gkogan@iki.rssi.ru} \email 2 {tsupko@iki.rssi.ru

$^3${Talk presented at the russian summer school-seminar "Modern
theoretical problems of gravitation and cosmology" \,
(GRACOS-2007), September 9-16, 2007, Kazan-Yalchik, Russia}}

\section{Introduction}

The possibility of formation of the primordial black holes (PBH)
with small masses at the early stages of Universe expansion was
discussed in papers \cite{ZN1}, \cite{CarrHawking}, see also
\cite{ZN2},\cite{ZN3},\cite{Khlop1},\cite{Khlop2}. Let us consider
the early stages of expansion, and PBH, massive enough for Hawking
temperature $T_H=\frac{\hbar c^3}{8\pi k GM}$ to be much less than
envi\-ron\-ment tem\-pe\-ra\-ture. In this case the main process
of interaction between PBH and environment is accretion of the
radia\-tion domina\-ted matter to PBH. In approximation of steady
flow, when the flow of mass $\dot M \equiv \frac{dM}{dt}$ is
calculated with formulae for stationary flow of gas, at rest in
infinity, to a gravitation center, the mass of PBH diverges if
initial time of accretion is close to beginning of Universe
expansion \cite{ZN1}. Authors \cite{ZN1} note that to answer the
question whether accretion to PBH is catastro\-phi\-cally high the
investigation of non-stationary problem is needed. Non-stationary
problem is solved numerically in paper \cite{PNN}.

Mentioned above papers consider non-rotating black holes defined
by Schwarzchild metric, and isotropic accretion of gas, which
increases PBH mass and keeps angular momentum equal zero. But it
is reasonable to assume formation of PBH with both mass and
angular momentum. The presence of rotation is a proper feature of
objects of different scales, beginning with ele\-men\-tary
particles (spin) and up to rotation of macro\-objects. As to PBH,
formation of PBH with both mass and angular momentum may result
from simultaneous action of potential and vortex initial
per\-tur\-bations.

It is shown for the first time by Doroshkevich \cite{dor66} that
due to effects of the general theory of relativity accretion of
particles (both non-relativis\-tic and relativis\-tic) to rotating
objects leads in general case to decrease of angular momentum of
the object. In papers \cite{Young1976},\cite{Young1977} accretion
of massive particles and photons to Kerr black hole is considered.
In particular, in paper \cite{Young1976} author discusses the
decrease of PBH angular momentum due to the isotropic accretion of
photons.

In present paper increase of mass and decrease of PBH angular
momentum are estimated taking into consideration non-stationarity
of the problem (see also \cite{BK&Ts}). We use results of
numerical solution of non-statio\-nary problem of PBH formation
and dynamics \cite{PNN} and approxi\-mate relations for accretion
into a rotating black hole \cite{Young1976}.

\section{Approximation of stationary accretion}

In approximation of stationary accretion one has the following
equation for increase the PBH mass \cite{ZN1}-\cite{ZN3}:
\begin{equation}
\label{dM/dt} \frac{dM}{dt} = \frac{27}{4} \pi R_G^2 c \rho_r \, ,
\end{equation}
where $M$ is black hole mass, $R_G = 2 G M/c^2$ is its
Schwarzchild radius, $\rho_r$ is the density of background matter.
At the radiation dominated stages the density in Universe falls as
$\rho_r = \frac{3}{32 \pi G t^2}$, what allows to integrate the
equation (\ref{dM/dt}). The integral depends weakly on upper bound
in time, therefore one may put it equal to infinity. Then we have:
\begin{equation}
\label{M(t)} M = \frac{M_0}{1 - \frac{81}{32} \frac{G M_0}{c^3
t_0}} \, , \label{mbh}
\end{equation}
where $M_0$ is the mass of black hole formed at the time $t_0$.
Formula (\ref{mbh}) diverges under $t_0 \rightarrow
\frac{81}{32}\frac{G M_0}{c^3} =\frac{81}{64}\frac{R_G}{c}$. Thus
if at the time of black hole formation $t_0$ its gravitational
radius $R_G$ was comparable with horizon radius $ct_0$, then in
stationary approximation accretion would lead to very high
increase of mass, as compared with initial. Note, that for $t \gg
\frac{81}{32} \frac{G M_0}{c^3}$ the growth of mass of PBH is
negligible, so for $M_{PBH} \sim 10^{15}$g we can completely
ignore it at $t \gg 10^{-20}$s from the Big Bang. Because of
decrease of PBH angular momen\-tum due to accretion, it would lead
us to conclusion about existence of almost non-rotating black
holes, even if they had considerable specific angular momentum
(non-dimensional) of the order of Kerr limit momentum
$a_{lim}=\frac{JG}{Mc^3}=1$ at the moment of formation. It is
clear that the latter conclusion is associated with assumption of
the stationary accretion and is not con\-vin\-cing. Rela\-tion
between the mass of just formed PBH and the Universe horizon,
which determines evolution of mass and angular momentum under the
subsequent accretion, can be found only by solution of
non-stationa\-ry problem.

\section{Estimations for increase of mass and
decrease of angular momentum}

As shown in previous section, formula (\ref{M(t)}) doesn't lead to
catastrophical accretion if the black hole mass is considerably
smaller than $c^3 t_0/G$. It means that black hole mass must be
smaller than the mass contained within Universe horizon at the
moment of time $t_0$. Non-stationary spherically symmetric problem
of PBH formation as a result of evolution of initial strong
pertur\-bations relative to Friedman background model is solved
numerically in paper \cite{PNN}. It is found that at the time of
its formation PBH will have mass considerably smaller than the
mass within the cosmological horizon, therefore catastrophical
accretion doesn't occur. This result is obtained for wide range of
initial conditions. It also agrees with the result of qualitative
analysis made in paper \cite{CarrHawking}.

The main idea of the present paper is to estimate the angular
momentum losses of PBH inherited at their birth, at the first
stages of the Universe expansion. For that purposes we combine the
results of the PBH mass evolution in hot Universe following from
the numerical simulations (solution of non-stationary problem) and
results for the angular momentum change due to accretion of the
photons. We use the results of numerical calculations \cite{PNN}
to estimate the mass growth and PBH angular momentum loss,
assuming the same rate of accretion. If the mass of formed PBH is
$M_0 = \eta \,\frac{32}{81} \frac{c^3 t_0}{G}$, where $\eta$ is a
non-dimensional coefficient, then in the absence of
catastro\-phi\-cal accretion ($\eta \ll 1$) we use the relation
for statio\-na\-ry accretion (\ref{mbh}) in order to estimate $M$,
and final mass of PBH is
\begin{equation}
\label{M} M = \frac{M_0}{1 -  \eta} \, .
\end{equation}

For estimation of $\eta$ we use results of solution of
non-stationary spherically symmetric problem obtained in paper
\cite{PNN}. As follows from the Table in that work, the mass $M_0$
of formed PBH doesn't exceed 10 per cent of horizon mass $M_{hor}$
at the time $t_0$ for all investigated in \cite{PNN} variants. The
coefficient $\eta$ is uniquely defined by relation $M_0/M_{hor}$:

\begin{equation}
 \rho_0=\frac{3}{32\pi G t_0^2},\,\,\,
 M_{hor}=\frac{4\pi}{3}\rho_0 (ct_0)^3=\frac{1}{8}\frac{c^3t_0}{G},\,\,\,
\end{equation}
\begin{equation}
 \eta =
\frac{81}{32}
\frac{GM_0}{c^3t_0}=\frac{81}{256}\frac{M_0}{M_{hor}}.
\end{equation}
Taking into account results \cite{PNN}, we obtain $\eta \approx
0.03$.

As shown in paper \cite{Young1976}, in case of accretion of the
isotropic photon flow (or other ultrarelativistic particles) to
rotating PBH, an evolution of specific angular momen\-tum
depending on mass can be written approximately in form $a \propto
M^{-10/3}$, i.e. $a=a_0(\frac{M_0}{M})^{10/3}$. Under $\eta \ll 1$
we obtain approximate dependences for increase of mass and
decrease of PBH angular momentum during the accretion:

\begin{equation}
M\approx M_0 (1+\eta),\,\,\, a\approx a_0
\left(1-\frac{10}{3}\eta\right).
\end{equation}
Under $\eta \sim 0.03$ we obtain that PBH mass $M$ increases by
$\sim 3$ per cent due to accretion, and its specific angular
momentum $a$ decrease by $\sim 10$ per cent.

Thus, PBH rapidly rotating at the time of formation will keep
their rapid rotation. If mass of PBH is not too large ($M \sim
10^{15}$ g) their evaporation becomes significant by the present
time \cite{fn98}. Ratio between diffe\-rent types of particles
generated during the evaporation depends essentially on angular
momentum of PBH \cite{fn98}, what influences on estimates of
different components of the background of modern Universe, such as
neutrino, gravitons, and other weakly interacting particles
\cite{bkr04}.

\Acknow {This work was partially supported by RFBR grant
05-02-17697, RAN Program "Formation and evolution of stars and
galaxies" \, and Grant for Leading Scientific Schools
NSh-10181.2006.2.}

\end{document}